\documentclass[11pt]{article}
\usepackage{graphicx}
\title{\bf One--Dimensional Primordial Gravitational Waves In Pure Quadratic Gravity}
\author{Taimur Mohammadi,\footnote{University of Applied Science and Technology, Kurdistan Branch , P.O.Box 66177-17792,
        Sanandaj, Iran. Email: t.mohammadi@uast.ac.ir}\\ Behrooz Malekolkalami,\footnote{Department of Physics,
University of Kurdistan, P.O.Box 66177-15175,
        Sanandaj, Iran. Email: b.malakolkalami@uok.ac.ir}\\
       }
\begin{document}
\date{\today}
\maketitle
\begin{abstract}
The almost scale--invariant spectrum for the stochastic background wave (in primordial Universe) is a firm prediction of inflationary scenarios. In the present work,
to study of primordial Gravitational  Waves,  one dimensional toy model in generalizations of the Einstein--Hilbert frame, described by second order curvature invariant
(called pure Quadratic Gravity)  is considered.  Solutions to the  primordial perturbations are more varied than Einstein--Hilbert  frame and include simple incoming and outcoming  waves. By examining the spectrum diagram of the perturbation solutions, it can be seen that the oscillating solutions are ruled out due to not having the necessary spectral power, but the solutions  as damping exponential have the ability to produce  a scale--invariant spectrum (confirmed by CMB data).
\end{abstract}

\medskip
{\small \noindent
\hspace{5mm}

\bigskip

\section{Introduction}
\indent
Gravitational Waves (\textbf{GW}) are disturbances propagated in spacetime which  have  different sources of production.  They were first predicted in $1916$ by  Einstein~\cite{GWs 1916}, and finally the first direct observation from merging massive black holes reported in  $2015$ (GW150914) and second in $2017$ (GW170817) by the Virgo collaboration and LIGO scientific collaboration ~\cite{Abbott 2016, Abbott 20171, Abbott 2017}. Undoubtedly, these observations promote the  knowledge of gravitational waves astronomy, leading to  understand and explore different and yet inaccessible phenomena in astrophysics and cosmology ~\cite{Saeed 2013}.
In high--energy astrophysics, GW are generated by  compact binaries systems whose dynamics is influenced by the dissipative effect of gravitational radiation--reaction forces. Sources such as
coalescing compact binary systems, stellar collapses and pulsars are all
possible candidates for detection ~\cite{Emmanuele 2022}. There are a number of efforts, worldwide, to detect gravitational wave, because, they can provide invaluable information on the geometry and content of our Universe as well as the nature of gravity ~\cite{Alice 2020,Emmanuele 2021}.\\
In major classification, the GW sources can be divided into two the following classes:\\
1) The relativistic astrophysical origin (as high energy events in extreme astrophysical environments including the gravitational wave bursts form instabilities in neutron stars or the gravitational wave signal from a supernova).
This class is usually output of events when matter is accelerated in an asymmetrical way,  but due to the nature of the gravitational interaction, significant levels of radiation are produced only when very a tremendous amount of mass is accelerated in very strong gravitational fields. Such an event cannot be simulated  or created in a
terrestrial experiment but is found in a variety of astrophysical systems where masses and movements are on the astronomical scale, as what happens in  supernova and binary systems.\\
2) The cosmological origin (as primordial GW produced in the early stages of the Universe, that is during the inflationary period and
reheating epochs). There could also be stochastic gravitational wave backgrounds  from the early universe, produced by non--equilibrium phenomena that carry a large amount of energy. Investigating the properties of these backgrounds
is an active area of research, which includes cases as fluctuations amplified during inflation, first--order phase transitions and cosmic strings ~\cite{Chiara 2018}.\\
Before general explanations about the two classes of sources mentioned, let us point out a fundamental difference between them. This difference  is related to their production mechanism and we will talk a little about it below.\\
The mechanism of producing the gravitational wave by classical sources (the first class)  is mainly caused by a time varying matter distribution with a non--zero quadrupole moment. GW are hence produced during many astrophysical phenomena that involve colliding or
collapsing bodies, such as binary star systems, coalescing black holes or supernova. Depending on the details of the system, the emitted GW from these point sources will peak at specific frequencies. The production mechanism of the first class is a classical
process and very different in nature to the second class which are originated from a stochastic background. This background did not originate from a source, but originated from quantum fluctuations in the spacetime metric. These (very early) vacuum fluctuations are assumed to be deviation from absolute homogeneity (small deviations away from the standard FRW model) and as the Universe expanded, some of these fluctuations were amplified\rlap.\footnote{one of the most firm predictions of all
inflationary models is the amplification of
vacuum fluctuations of the metric tensor and the formation of a primordial,
stochastic background of relic GW.}
These fluctuations were stretched to superhorizon scales during the inflationary phase and result in a scale invariant spectrum of tensor fluctuations.\\
one of the most extreme events known to astronomy (for the first class) is gravitational collapse. A supernova occurs during the last evolutionary stages of a massive star (the gravitational collapse of the interior degenerate core of an evolved star) and  the expected  result is a neutron star or black hole. The collapse releases an enormous amount of energy (about 1046 joules), most of which is carried away by neutrinos, an uncertain fraction is converted into GW ~\cite{Chiara 2018, Vittorio 2023, Falco 2024, Battist 2022, Battist 2023}.

Also, one  fact is that compact objects offer the ultimate strong field tests of general relativity through the gravitational radiation emitted when black holes form. Or, the binary pulsars merging occurs  under the effect of gravitational wave energy losses, that is the binary system emits an immense  amount energy  in the form of GW as the orbit shrinks until the two objects merge into a single  spinning black hole. These waves  carry  information about the black hole and the merging event which should be detected by
the gravitational wave observatory instruments. The first detector (used in  a gravitational wave observatory) was made in September 2015 by the Advanced LIGO observatories, detecting gravitational waves with wavelengths of a few thousand kilometers from a merging binary of stellar black holes.  LIGO is sensitive to gravitational waves only in the frequency range from about 5 Hz to about 20,000 Hz. Fortunately, this range includes the gravitational wave frequencies  emitted from type II supernovas and  the merger of a pair of neutron stars. There are several other ground--based projects  are designed to detect GW at completely different frequencies\rlap.\footnote{Examples of these projects are:  Japanese TAMA and French--Italian VIRGO projects.}
The frequency range of a gravitational wave signal provides information about its source: the lower the frequency, the larger the mass involved. It also tells scientists which type of detector to use to look for which source, as the detector size should be comparable to the wavelength of the signal. In addition to astronomical phenomena such as  supernova and black holes, gravitational wave detectors also allow astronomers to take a closer look at cosmological phenomena as the early universe events.\\
The standard cosmology is based on the idea that the universe has reasonably homogeneous and isotropic structure (in the large scale). The observational evidences indicate that Universe is  in an acceleration phase today and  then to describe a relatively complete evolving universe, the descriptive equations must include parameters such as the composition of the universe, its current expansion rate and the initial spectrum of density perturbations. The initial scalar density perturbations which  have been generated during inflation are assumed to be the origin of  large scale structure  observable today.  But scalar density perturbations are not the only perturbations, since the inflationary theory suggests that  tensor perturbations have also  been generated  as source of gravitational radiation so--called primordial GW (\textbf{PGW}). In other words, during  cosmic inflation, PGW have also been  generated through the same physical process that seeds all structure formation in the observable universe. They have been propagating in spacetime  since the inflationary period and have not been detected to date. It is hoped that they could be detected by the next generations of more sensitive instruments if their amplitude is large enough.

Detection of PGW would have a profound impact on revealing the mysteries of the very early universe as  the vacuum fluctuations are  amplified through  inflationary period. Furthermore, the relic gravitational wave can  give information on the state of the very early universe
and the corresponding energy scales, because the power spectra  of PGW  is a direct measure of the expansion rate of the Universe at the time that wavelength was stretched beyond the horizon. In other words, the spectrum of energy carried by PGW can represent a measure of the energy of that period and challenge the ability of a direct detection of the relic waves by the current gravitational detector.  Although current technical knowledge is not able to detect PGW, there are other ways to reveal the traces of them as  CMB anisotropic. For example, the LISA is a complementary project to CMB experiments which aims to  detect and accurately measure gravitational wave by using laser interferometry. It is believed that both electric and magnetic polarization of  CMB   are generated  by gravitational wave, specially, the detection of the B--mode signal from stochastic background gravitational wave is one of the important goals of these studies.~\cite{Falco 2024, Durmus 2023, Oikonomou 2022,Rong 2017,  Ji 2023, Zhe 2023}.

Generally, spectral methods are a class of techniques used in applied mathematics and scientific computing to numerically solve certain differential equations.
These methods are used in both linear and nonlinear wave propagation problems, specially when dealing with stochastic waves.   In this paper, we utilize  this method to analysis the PGW. It is useful to express any gravitational wave estimates in terms of the corresponding energy spectrum given by:
\begin{equation}\label{N7}
\Omega_{h}(k, t)=\frac{1}{\rho_c}\frac{d\rho}{d\ln k },
\end{equation}
where $\rho$  and $\rho_c$ are gravitational wave energy density and critical energy density, respectively. More precisely, it characterizes
the spectrum of the relic gravitational radiation present today ($t=t_0$) inside our horizon, and thus accessible to direct observations.\\
As mentioned above, we use the spectral method to analysis the PGW. However, the corresponding differential equations is derived by considering perturbations in an extended theory of gravity, which is discussed below.\\
In the context of gravitational wave astronomy, the GW are employed to inquire gravitational phenomena either within General Relativity (GR) or alternative frameworks (as extended gravitation theories), as well as  primordial processes occurring in the early universe shortly after the Big Bang.~\cite{Falco 2024}.

The study of the PGW and their spectrum discussed above has been done in the framework of the GR in the literature.  GR has a very large area of applications (confirmed experimentally), ranging from the solar system  tests to the dynamical phenomena, like accretion processes by compact objects, involving the flow of gas streams around massive stars. However, these astrophysical phenomena  are restricted to a small scale, of the order of the radius of compact objects, black holes, or of the solar system.
Means that, GR faces conceptual and observational challenges when applied to the large scales, since the gravitational force dominates not only in
solar system, but extends far beyond in the realm of galaxies, and of the entire Universe. Many astrophysical phenomena (as PGW and gravitational redshift)
are of this type that are belong to  strong gravity regime and high energy events ~\cite{Vittorio 2023, Noah 2016, Zack 2020}. Therefore, the study of these phenomena in Modified Gravity (MG) theories may  provide some specific signatures and effects, which could distinguish and discriminate between the various theories of gravity. For example, the strong constraint on the speed of GW ruled out many classes of MG models.

Based on this motivation, studying the  gravitational phenomena  in MG Theories can be justified and attractive as such considering the gravitational wave produced in inflation scenarios \cite{Guzzetti 2016}, cosmic expansion at very early universe and the
current period \cite{ Nojiri 2018}, or  Cosmic Acceleration \cite{MARK 2007, Andrzej 2012} and Cosmological Constant Problem, based on MG theories  \cite{Clifton 2012, Jibril 2024}.\\
In the present work, we  focus on  the  power spectrum of PGW through a one--dimension (toy) model where the gravity frame is described by the pure quadratic (Curvature Scalar)gravity ($f(R)=R^{2}$)~\cite{ Tomohiro 2020}\rlap.\footnote{One of the well--known of the quadratic action is Starobinsky model for Cosmic Inflation. Other kinds of $f(R)$ models can be found in~\cite{Soham 2018, Naf 2011, Sotiriou 2010}} \\
To better explain the motivation of the one--dimensional model,
we note that the inflationary theory   predicts an epoch in the early  Universe when space expansion is governed by a large factor. Since this expansion was not symmetric spatially, it may have emitted gravitational radiation  detectable today as a (tensor perturbations) gravitational  wave background. The spectrum of this perturbations has a nearly scale--invariant  pattern. This background signal is too weak for any currently  gravitational  wave interferometers to observe, however its detection would open a new window to understanding  the origin and evolution of the very early Universe.\\ One motivation to take a one dimensional model is that,  the  spectrum of density perturbations generated has a pattern similar to scale--invariant  spectrum of cosmological perturbations generated by cosmic inflation. The next motivation in choosing this one dimensional model can be an inspiration to analysis and solve  two or three dimensional problems. In addition, one of the other advantages of this model can be mentioned is the variety of exact solutions of the corresponding differential equation.

The work is organized as follows: In section 2, we obtain the motion equations for gravitational perturbations in quadratic gravity for an isotropic and perfect fluid. In section 3 we introduce  and discuss the one--dimensional model and the obtain spectrum of the perturbations. Conclusions are given in section 4.
\section{Evolution of Perturbations  in f(R) Gravity Formalism}
In order to obtain the motion equations for gravitational perturbations, we consider a perturbed Friedman--Robertson--Walker (FRW) metric with the line element shown as~\cite{Beyond 20018}
\begin{equation}\label{N7}
ds^2=g_{\mu\nu}dx^{\mu}dx^{\nu}=\left(\bar{g}_{\mu\nu}+h_{\mu\nu}\right)dx^{\mu}dx^{\nu} = a^{2}\left[-d\tau^2+ \left(\delta_{ij}+h_{ij}(t,\textbf{x})\right)dx^i dx^j\right]
\end{equation}
where $\tau$ is conformal time,  $\bar{g}_{\mu\nu}= diag\{-a^{2}, a^{2},a^{2},a^{2}\}$ is the unperturbed FRW background  and $h_{\mu\nu}$ are the perturbations  satisfying the conditions: symmetric ($h_{ij}$=$h_{ji}$), traceless ($h^i_{i} = 0)$ and transverse $(h^j_{i,j} = 0)$ and also $|h_{\mu\nu}|<<1$,\hspace{2mm}$h_{00}=h_{0i}=0$.\\
To write down the perturbational  equations, we use  the lagrangian formalism.  For  an isotropic and perfect fluid, we have $h_{ij}=h_{ji} $\hspace{1mm} and  the anisotropic stress of the energy--momentum tensor~\cite{Steven 2003} is zero. Hence, the motion equations take the following form~\cite{Taimur 2020}
\begin{equation}\label{C00}
\partial_{\mu}\left(\sqrt{-\bar{g}} \frac{\partial  f(R)}{\partial(\partial_{\mu}h_{ij})}\right)=0,
\end{equation}
which by using the chain rule becomes
\begin{equation}\label{C00}
\partial_{\mu}\left(\sqrt{-\bar{g}} \frac{df}{dR} \frac{\partial R}{\partial(\partial_{\mu}h_{ij})}\right)=0,
\end{equation}
where $\bar{g}$ is the determinant of $\bar{g}_{\mu\nu}$ and Ricci scalar is given by~\cite{Beyond 20018, Maurizio 2017}.
\begin{equation}R=\left(\frac{\bar{-g^{\mu\nu}}}{64\pi G } \partial_{\mu}h_{ij}\partial_{\nu}h_{ij}\right).\end{equation}
By substituting  $f(R)=R^{2}$ into (4) and using the isotropic condition $h_{ij}(\tau,\textbf{x})=h(\tau,\textbf{x})$, it reduces to
\begin{equation}\label{C00}
\partial_{\mu}\left(\sqrt{-\bar{g}}\hspace{1mm} R \hspace{1mm}\bar{g}^{\mu\nu} \partial_{\nu}h\right)=0.
\end{equation}
This equation is the motion equation for the gravitational  perturbations in quadratic gravity and in the next section, we treat it for propagation of perturbations in one dimension, say $x$.
\section{One--Dimensional Solutions}
By considering the  propagation of perturbations  in $x$ direction, that is $h(\tau,\textbf{x})=h(\tau,x)$, equation (6) reads
\begin{equation}\label{C00}
\partial_{\tau}\left(\sqrt{-\bar{g}}\hspace{1mm} R \hspace{1.5mm}\bar{g}^{00}\hspace{1mm} \partial_{\tau}h\right)+\partial_{x}\left(\sqrt{-\bar{g}}\hspace{1mm} R \hspace{1.5mm}\bar{g}^{11} \hspace{1mm}\partial_{x}h\right)=0,
\end{equation}
and also Ricci scalar (5) takes the form $R=\frac{1}{64\pi G a^{2}}\left((\partial_{\tau}h)^{2}-(\partial_{x}h)^{2}\right)$. After substituting this scalar into (7) and simplification, we find
\begin{equation}\label{C00}
-\partial_{\tau}\left(\left[(\partial_{\tau}h)^{2}-(\partial_{x}h)^{2})\right]\partial_{\tau}h\right)+\partial_{x}\left(\left[(\partial_{\tau}h)^{2}-(\partial_{x}h)^{2})\right]\partial_{x}h\right)=0,
\end{equation}
this  is a nonlinear equation and as we know nonlinearity can have a different nature to the equations and their solutions. Unlike linear equations, there are much less literature exists on the solution of nonlinear partial differential equations. Thus, depending on form of the equation, methods to obtain solutions can be different. For example, one of the useful technics is to use the symmetric form of the equation, from which the solution can  be predicted. On the other hand, solving equations by computer algebra methods has also attracted many people. Fortunately, for equation (8), both symmetrical and computer algebra  methods lead to the following general solutions:
\[h(\tau,x) =  \left\{\begin{array}{lr}
f(x+\tau),&\\
g(x-\tau),
\end{array}
\right.\]
where $f, g$ are arbitrary functions. Note that due to the nonlinearity, the linear combination can't be a solution. The above general solutions includes also the familiar plane wave
\begin{equation}\label{C00}
e^{i(x \pm \tau)},
\end{equation}
therefore, in such universe, the plane waves can be only in the incoming or outcoming form. Anyway, such waves do not have the required  power to detect today and this can be verified by the spectral power formula for perturbations in (FRW) expanding universe, that is \cite{Yuki 2006}
\begin{equation}\label{C00}
\Omega_{h}(\tau,k)=\frac{\Delta_{h,prim}^{2}}{12 a^{2}(\tau)H^{2}(\tau)}[\partial_{\tau}T(\tau,k)]^{2},
\end{equation}
where $\emph{T}(\tau,k)$ is the transfer function and  $\Delta_{h,prim}^{2}=\frac{16}{\pi}\left(\frac{H_{inf}}{m_{Pl}}\right)^{2}$ \footnote{$H_{inf}$
is the Hubble constant during inflation and $m_{Pl}$ is the Planck Mass.}\cite{Yuki 2006, Bernal 2019}\rlap.\hspace{1mm}  So, to detect the PGW in the present age, one have to look for  solutions that produce a proper spectral power. For example, we consider the following damped perturbation
\begin{equation}h(\tau,x)=a\exp(-b |x-\tau|) \hspace{3mm}  \hspace{3mm} a,b >0,
\end{equation}
The Fourier transform of $h(\tau,x)$ is:
\begin{equation}h(\tau,k)=a\exp(-i k \tau) \hspace{3mm}\frac{\sqrt{2}}{\sqrt{\pi}}\frac{b}{b^{2}+ k^{2}} ,\end{equation}
and use the equation (10) to calculate its spectral power. The graph for the spectrum of  perturbation (11) is illustrated in Fig.1 in which $\Omega(k)= \Omega_{h}(\tau_0,k)$, $a(\tau_0)=1$ and we assume $a= b =3$. To be more precise, Fig.1 shows the spectrum of the (tensorial) primordial perturbations produced in the geometry of the very early universe, at the present time $\tau=\tau_0$, in terms of the comoving wave number. In the presented frequency range, it is seen that by increasing the wave number, the spectrum takes  a scale--invariant character. Such a character for  primordial fluctuations is predicted   by the cosmic inflationary scenarios based on the symmetry breaking phase transition of a self--ordering scalar field~\cite{Smith 2008}.\\
In the end, we draw the reader's attention to two points about the spectral graphs show in Fig.1 and Fig.2 :\\
1) By increasing  numerical coefficients $a, b $, the graph moves to the right and up (and vise versa).\\
2) These  graphs  qualitatively resemble the spectral graph for a driven harmonic  oscillator where the  main similarities return to  the presence of a spectrum peak and decreasing  character at high frequencies.
\begin{figure}[!tt]
\centering
 \includegraphics[width=0.6\textwidth]{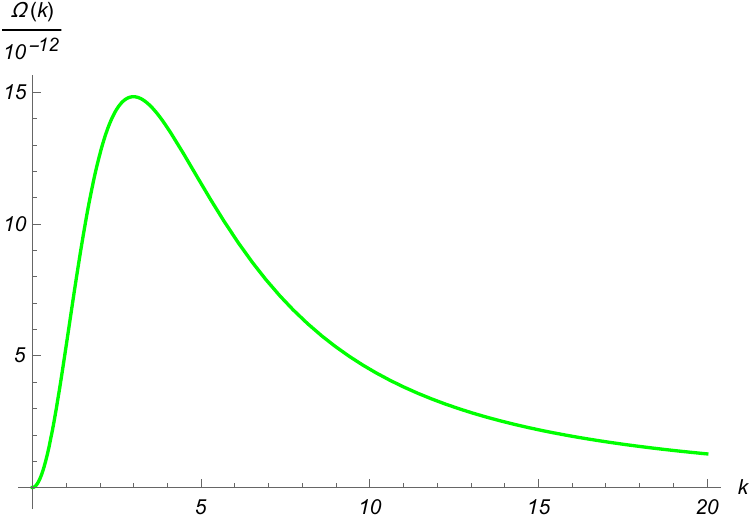}
 \caption{The power spectrum $\Omega_{h}( k)$ with respect to $k$ }.
  \label{5bl0tn}
\end{figure}

\begin{figure}[!tt]
\centering
 \includegraphics[width=0.6\textwidth]{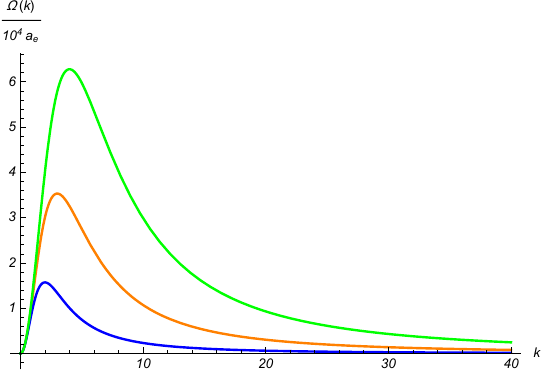}
 \caption{The power spectrum $\Omega_{h}( k)$. Blue: $a=b=2$, Orange: $a=b=3$ and Green: $a=b=4$ }.
  \label{5bl0tn}
\end{figure}
\section{Conclusions}
Inflationary models typically predict the existence of  (very early) tensorial perturbations leading to  relic GW background. It is true that they have not been revealed until today,  this could be tested in a close future by experiments either indirectly (as, considering that it an imprint on CMB B--mode polarisation anisotropic), or even directly with gravitational waves interferometers such as LIGO and VIRGO. Also,  an almost scale--invariant power spectrum of primordial fluctuations, in a spatially flat Universe is a natural outcome of the  inflationary scenario in the early Universe (confirmed by CMB data).

For this purpose and to find solutions to primordial  perturbations, we considered a one dimensional (toy) model in pure quadratic (in  curvature invariant) modified gravity.
The non--linearity of motion equations allows  the variety of  solutions. The use of the spectral method indicates that there is no chance to detect for Harmonic perturbations, that is, they  are not able to produce the desired spectrum . But instead, a damping exponential perturbations can produce the desired  spectrum, that is,
a scale--invariant spectrum in high frequency regime.

As last  point, we note that   the spectrum graph of primordial perturbations (in this one dimensional model)  is  qualitative similarity with that of the driven harmonic oscillator. This point is important from the point of view that in many classical and quantum systems, the harmonic oscillator provides an excellent description of periodic systems such as a mass on a spring, resonating electronic LC circuits (in classical regime) or a number  of bosonic systems and phonons in an elastic solid  (in quantum realm). Therefore, perhaps the conclusion that, some phenomena in the very early Universe (as propagating stochastic background wave) can be described by driven harmonic oscillator, is not far from the truth.

\end{document}